\documentclass[aps,prd,groupedaddress,showpacs,superscriptaddress,eqsecnum,floatfix,twocolumn]{revtex4-1}
\usepackage{graphicx}
\usepackage{dcolumn}
\usepackage{bm}
\begin{document}

\title {Shape and dynamics of nonrelativistic vortex strings in parity-breaking media.}

\author{A.~A.~Kozhevnikov}
\email[]{kozhev@math.nsc.ru} \affiliation{Laboratory of
Theoretical Physics, S.~L.~Sobolev Institute of Mathematics, Novosibirsk, Russian
Federation}
\affiliation{Novosibirsk State University, Novosibirsk, Russian
Federation}

\date{\today}

\begin{abstract}
The  shape and dynamics of the nonrelativistic gauge vortex string  in the  parity-broken media is considered, upon reducing the problem to finding the extremum of the  Abelian Higgs model effective action with the fixed B-type helicity of the gauge field. It is shown that in contrast with the case of the fixed A-type helicity,  the static solution of the Ginzburg-Landau energy functional in the London limit is the helix with the specific relation between the curvature and torsion of the vortex line depending on the strength of the space parity violating contribution of the Lifshitz invariant.  A nonlinear dynamical equation is linearized in case of small oscillations around helical contour, and the polarization and dispersion law of the propagated waves are obtained.
\end{abstract}

\maketitle

\section{Introduction}
~

The interest in physical effects which could take place in the intrinsically space parity non-symmetric media is twofold. First, such type of the environment  can be realized in heavy ion collisions where the configuration of the electromagnetic fields, due to the macroscopic manifestation of the quantum axial anomaly \cite{abj1,abj2}, results in appearance of the current along the direction of magnetic field. The physical manifestation of such a current is the chiral magnetic effect (CME). See Refs.~\cite{kharz08,fukush08} and Ref.~\cite{kharz14} for a review. At the level  of the effective Lagrangian density such effect is described by the inclusion of the term $\propto({\bm A}\cdot[{\bm\nabla}\times{\bm A}])$, with ${\bm A}$ being the vector potential of the gauge field. The second possibility refers to the  unconventional superconductivity in the crystals without inversion center where the Ginzburg-Landau energy  functional includes additional terms, the so called Lifshitz invariants \cite{lifsh}, whose form, in particular, corresponds to the scalar product of the current and the strength ${\bm B}$ of the magnetic field $\propto\int d^3x({\bm j}\cdot{\bm B})\sim\int d^3x({\bm B}\cdot[{\bm\nabla}\times{\bm B}])$ \cite{levit85,mineev94,samokhin04,kaur05}. In a more formal context, the inclusion of the above mentioned terms follows from fixing the averaged value of the $P$-odd $A$-type helicity,
\begin{equation}\label{haaver}
\langle h_A\rangle=\frac{1}{t_{fi}}\int_{t_i}^{t_f}dt\int d^3x({\bm A}\cdot[{\bm\nabla}\times{\bm A}]),
\end{equation} in the Feynman path integral for the gauge field,
\begin{widetext}
\begin{eqnarray}\label{Kfi}
K_{\langle h_A\rangle {\rm fixed}}&=&\int D[A]D[\psi_M]e^{iS[A,\psi_M]}\delta\left(\frac{1}{t_{fi}}\int d^4x({\bm A}\cdot[{\bm\nabla}\times{\bm A}])-\langle h_A\rangle\right),
\end{eqnarray}\end{widetext}
where $\psi_M$ stands for some matter fields irrelevant for the present qualitative discussion. Using the exponential representation for $\delta$-function one arrives at the mentioned terms in the effective action of the gauge fields. Correspondingly, one can study the Feynman path integral for the gauge field configurations with the fixed average $B$-type helicity obtained from Eq.~(\ref{haaver}) by the replacement ${\bm A}\to{\bm B}=[{\bm\nabla}\times{\bm A}]$,
\begin{equation}\label{hB}
h_B=\int d^3x({\bm B}\cdot[{\bm\nabla}\times{\bm B}]).
\end{equation}
This type of helicity was proposed, for example, in Ref.~\cite{zeld}. In both mentioned cases, the search for the saddle point solutions for the semiclassical evaluation of the path integral would reduce to minimization of the effective action which include the parity-odd contribution.

Recent works \cite{kharz20,babaev20} deal with the configuration of the magnetic field strength of the vortices in the non-centrosymmetric superconductors where the Ginzburg-Landau energy functional includes the Lifshitz invariant of the form $({\bm j}\cdot{\bm B})$ corresponding to the effective action with fixed B-type helicity. The case of the straight static vortices was considered in these papers.

The purpose of the present work is to consider the case of the curved Abrikosov-Nielsen-Olesen (ANO) gauge string vortices \cite{abr,niel} in the situation of  environment with the explicitly  broken space parity analogous to that studied in Refs.~\cite{kharz20,babaev20}. It is well-known, that the Ginzburg-Landau free energy functional describing the Abrikosov magnetic vortices in the type II superconductors in the limit of vanishing temperature \cite{abrikosov87,kleinert}, is equivalent to the nonrelativistic limit of the Abelian Higgs Model (AHM). By this reason the time-dependent variant of AHM will be taken as the basis of the following treatment of the vortex shape and its dynamics in the media with explicit space parity breaking. In its application in the condensed matter physics (particle physics), AHM is characterized by two length (mass) parameters, namely, the London penetration depth $\lambda_L$ (the mass of the gauge boson $m_V$) and the correlation length $\xi$ (the Higgs mass $m_H$). The following treatment will assume  the London limit $\ln\lambda_L/\xi\gg1$ ($\ln m_H/m_V\gg1$). In this limit, the results can be obtained in analytic form.

The subsequent material is organized as follows. Section \ref{sec1} is devoted to the derivation of the energy functional for nonrelativistic curved ANO gauge string vortex with the parity-odd contribution  and its variation over the contour shape. The equation of motion and its static solution are considered in Sec.~\ref{eqmo}. Sec.~\ref{oscill} contains the study of the small oscillations around the helical contour shape including the dispersion law of the waves and their polarization properties. The discussion of the obtained results are given in Sec.~\ref{discus}. As for the notations,  we keep the velocity of light $c$ and Planck constant $\hbar$ in all formulas throughout the text.

\section{The effective action}
\label{sec1}
~

The starting point is the effective action of the time-dependent nonrelativistic AHM with the gauge vortices \cite{kozh10,kozh15}:
\begin{eqnarray}
S&=&\int d^4x\left\{-\frac{1}{8\pi}{\bm B}^2-\frac{g}{2}(|\psi|^2-n_0)^2+\right.\nonumber\\&&\left.\frac{1}{2}[\psi^\ast(i\hbar\partial_t+qa_0)\psi
+{\rm
c.c.}]-\right.\nonumber\\&&\left.\frac{1}{2m}\left|\left(-i\hbar{\bm\nabla}-\frac{q}{c}{\bm
A}+\frac{q}{c}{\bm a}\right)\psi\right|^2\right\}.
\label{S}\end{eqnarray}As is discussed in Refs.~\cite{kozh10,kozh15}, the contribution from the electric field in the charge-neutral environment is suppressed by the square of the velocity of light. Let us write this action in the London limit characterized by the assumption of the constant density of the condensate, $\psi=n_0^{1/2}e^{i\chi_{\rm reg}}$ everywhere except the core of the vortex line where it vanishes at the distance $\simeq\xi$. Here, the regular phase $\chi_{\rm reg}$ can be set to zero while the singular phase $\chi_s$ responsible for the vortex \cite{abrikosov87,kleinert}, see below, is already included via the terms with $a_0$ and ${\bm a}$ in Eq.~(\ref{S}).  Adding  the parity-breaking term ($\propto\gamma$) introduced in refs.~\cite{kharz20,babaev20}, one obtains the expression
\begin{equation}\label{Seff}
S_{\rm eff}=\int dtd^3x\left(n_0qa_0-\mathcal{E}\right),
\end{equation}where
\begin{eqnarray}\label{F}
\mathcal{E}&=&\frac{1}{8\pi}\left[{\bm B}^2+\frac{1}{\lambda^2_L}({\bm A}-{\bm a})^2-\frac{2\gamma c}{\lambda^2_L}{\bm B}({\bm A}-{\bm a})\right].
\end{eqnarray}Since the charged current ${\bm j}$ in presence of the vortex is given by Eq.~({\ref j}) below, the parity-breaking term in Eq.~(\ref{Seff}) looks like ${\bm B}\cdot{\bm j}\propto{\bm B}\cdot[{\bm\nabla}\times{\bm B}]$ and corresponds to fixing the B-type helicity in the energy functional. As for the notations, $n_0$ is the density of condensate, $q$ ($m$) is the charge (mass) of the scalar field particle, $c$ is the velocity of light, and $$\lambda_L=\sqrt{\frac{mc^2}{4\pi n_0q^2}}$$ is the London penetration depth.   The four-vector $a_\mu=-\frac{\hbar c}{q}\partial_\mu\chi_s$ proportional to the gradient of the singular phase $\chi_s=\chi_s(t,{\bm x})$, of the scalar field,
$$[{\bm\nabla}\times{\bm\nabla}]\chi_{\rm s}=2\pi\int d\sigma{\bm
X}^\prime\delta^{(3)}({\bm x}-{\bm
X(\sigma,t)}),$$
describes the presence of gauge vortex  whose shape in its dependence on time is given by ${\bm X}(\sigma,t)$. Explicit expressions for the space Fourier components of $(a_0,{\bm a})$ are \cite{kozh10}
\begin{eqnarray}\label{amuk}
a_{0{\bm k}}&=&-i\frac{\Phi_0}{c{\bm k}^2}\int d\sigma{\bm k}\cdot\left[\dot{{\bm X}}\times{\bm X}^\prime\right]e^{-i{\bm k}\cdot{\bm X}},\nonumber\\
{\bm a}_{\bm k}&=&i\frac{\Phi_0}{{\bm k}^2}\int d\sigma\left[{\bm k}\times{\bm X}^\prime\right]e^{-i{\bm k}\cdot{\bm X}}.
\end{eqnarray}Hereafter, the overdot (prime) stands for the derivative over time (the length parameter $\sigma$) of the contour variable ${\bm X}\equiv{\bm X}(\sigma,t)$; $\Phi_0=2\pi\hbar c/q$ is the flux quantum. The vortex with single quantum of magnetic flux is assumed in the present treatment. Because the current in this model is
\begin{eqnarray}\label{j}
{\bm j}&=&\frac{c}{4\pi\lambda_L^2}({\bm a}-{\bm A}),
\end{eqnarray}the term $\propto\gamma{\bm j}\cdot{\bm B}$ in Eq.~(\ref{Seff}) is the example of the Lifshitz invariant which  violates space parity \cite{levit85,mineev94,samokhin04,kaur05,kharz20,babaev20}. The energy $E=\int d^3x\mathcal{E}$ is minimized by the field strength ${\bm B}$ which can be found from equation similar, up to replacement $c\to 1/k$, with that from Refs.~\cite{kharz20,babaev20}:
\begin{eqnarray}\label{eqforB}
{\bm\nabla}\times{\bm B}+4\pi\gamma{\bm\nabla}\times{\bm j}&=&\frac{4\pi}{c}{\bm j}+\frac{\gamma c}{\lambda^2_L}{\bm B}.
\end{eqnarray}The space Fourier components of ${\bm A}_{\bm k}-{\bm a}_{\bm k}$ and ${\bm B}_{\bm k}$ found from Eq.~(\ref{eqforB}) are
\begin{eqnarray}\label{AB}
{\bm a}_{\bm k}-{\bm A}_{\bm k}&=&\left[\left({\bm k}^2+\frac{1}{\lambda^2_L}\right)^2-\frac{4\alpha^2}{\lambda^2_L}{\bm k}^2\right]^{-1}\times\nonumber\\&&\left[{\bm k}^2\left({\bm k}^2+\frac{1-2\alpha^2}{\lambda^2_L}\right){\bm a}_{\bm k}+\right.\nonumber\\&&\left.i\frac{\alpha}{\lambda_L}\left({\bm k}^2-\frac{1}{\lambda^2_L}\right)[{\bm k}\times{\bm a}_{\bm k}]\right],\nonumber\\
{\bm B}_{\bm k}&=&\left[\left({\bm k}^2+\frac{1}{\lambda^2_L}\right)^2-\frac{4\alpha^2}{\lambda^2_L}{\bm k}^2\right]^{-1}\times\nonumber\\&&\left\{i\frac{[{\bm k}\times{\bm a}_{\bm k}]}{\lambda^2_L}\left({\bm k}^2+\frac{1}{\lambda^2_L}-2\alpha^2{\bm k}^2\right)-\right.\nonumber\\&&\left.\frac{\alpha}{\lambda_L}\left({\bm k}^2-\frac{1}{\lambda^2_L}\right){\bm k}^2{\bm a}_{\bm k}\right\}.
\end{eqnarray}Hereafter we introduce the new dimensionless parameter which characterizes the parity-breaking effects,
\begin{equation}\label{alpha}
\alpha=\frac{\gamma c}{\lambda_L}.
\end{equation}The energy is represented in the form
\begin{eqnarray}\label{Ftot}
E&=&\frac{1}{8\pi}\int\frac{d^3k}{(2\pi)^3}\left\{\frac{{\bm k}^2}{\lambda^2_L}\left({\bm k}^2+\frac{1}{\lambda^2_L}\right)|{\bm a}_{\bm k}|^2\times\right.\nonumber\\&&\left.\left(1-\alpha^2\right)+2i\alpha(1-\alpha^2)\frac{{\bm k}^2}{\lambda^3_L}({\bm k}\cdot[{\bm a}_{\bm k}\times{\bm a}^\ast_{\bm k}])\right\}\times\nonumber\\&&\left[\left({\bm k}^2+\frac{1}{\lambda^2_L}\right)^2-\frac{4\alpha^2}{\lambda^2_L}{\bm k}^2\right]^{-1}.
\end{eqnarray}
The gauge vortex contour shape ${\bm X}$ will appear through the following quantities:
\begin{eqnarray}\label{shape}
|{\bm a}_{\bm k}|^2&=&\frac{\Phi^2_0}{{\bm k}^2}\int d\sigma_1d\sigma_2({\bm X}^\prime_1\cdot{\bm X}^\prime_2)e^{i{\bm k}\cdot{\bm X}_{21}},\nonumber\\
({\bm k}\cdot[{\bm a}_{\bm k}\times{\bm a}^\ast_{\bm k}])&=&\frac{\Phi^2_0}{{\bm k}^2}\int d\sigma_1d\sigma_2({\bm k}\cdot[{\bm X}^\prime_1\times{\bm X}^\prime_2])\times\nonumber\\&&e^{i{\bm k}\cdot{\bm X}_{21}},
\end{eqnarray}where ${\bm X}_{21}={\bm X}_2-{\bm X}_1$ and ${\bm X}_{2,1}\equiv{\bm X}(\sigma_{2,1},t)$. The time variable will be omitted in what follows. The integration over wave vector ${\bm k}$ can be made analytically to obtain
\begin{eqnarray}\label{Ffin}
E&=&\frac{\Phi^2_0\left(1-\alpha^2\right)}{8\pi\lambda^2_L}\int d\sigma_1d\sigma_2\left\{I_1(|{\bm X}_{21}|)({\bm X}^\prime_1\cdot{\bm X}^\prime_2)+\right.\nonumber\\&&\left.\frac{2\alpha}{\lambda_L}\left([{\bm X}^\prime_1\times{\bm X}^\prime_2]\cdot\frac{\partial}{\partial{\bm X}_{21}}\right)I_2(|{\bm X}_{21}|)\right\},
\end{eqnarray}where
\begin{eqnarray}\label{I12}
I_1&=&\frac{e^{-|{\bm X}_{21}|\sqrt{1-\alpha^2}/\lambda_L}}{4\pi|{\bm X}_{21}|}\left(\cos\frac{\alpha|{\bm X}_{21}|}{\lambda_L}+\right.\nonumber\\&&\left.\frac{\alpha}{\sqrt{1-\alpha^2}}\sin\frac{\alpha|{\bm X}_{21}|}{\lambda_L}\right),\nonumber\\
I_2&=&\frac{\lambda^2_Le^{-|{\bm X}_{21}|\sqrt{1-\alpha^2}/\lambda_L}}{8\pi\alpha\sqrt{1-\alpha^2}|{\bm X}_{21}|}\sin\frac{\alpha|{\bm X}_{21}|}{\lambda_L}.
\end{eqnarray}The expression for $E\equiv E[{\bm X}]$ is nonlocal, but the scale of the non-locality, at $\alpha$ not too close to 1, is $\lambda_L$, so one can hope to reduce the expression to the local form in the London limit, where $\lambda_L$ is relatively small. The criterion of this will be established below. As is pointed out in Ref.~\cite{kharz20}, the condition of stability (the positivity of free energy) requires $\alpha\leq1$. In physical terms, the effective penetration length $\lambda_{\rm eff}=\lambda_L/\sqrt{1-\alpha^2}\to\infty$  at $\alpha\to1$ which means that the magnetic flux is not confined inside the tube of a finite transverse size, and the magnetic field spreads everywhere.

When obtaining the equation for the determination of the vortex shape ${\bm X}(\sigma,t)$ one should vary the effective action Eq.~(\ref{Seff}). However, Eq.~(\ref{Ffin}) is not very suitable for this purpose. A convenient way is to vary over ${\bm X}$   before the integration over wave vector ${\bm k}$. When so doing, one should require the condition ${\bm X}(\sigma_f,t)={\bm X}(\sigma_i,t)$, in order to drop the surface terms when integrating over $\sigma$ by parts. Then one finds that
\begin{widetext}
\begin{eqnarray}\label{variate}
\delta\int d\sigma_1d\sigma_2({\bm X}^\prime_1\cdot{\bm X}^\prime_2)e^{i{\bm k}\cdot{\bm X}_{21}}&=&i{\bm k}\cdot\int d\sigma_1d\sigma_2\left([{\bm X}^\prime_2[{\bm X}^\prime_1\times\delta{\bm X}_1]]-[{\bm X}^\prime_1[{\bm X}^\prime_2\times\delta{\bm X}_2]]\right)e^{i{\bm k}\cdot{\bm X}_{21}},\nonumber\\
\delta\int d\sigma_1d\sigma_2i({\bm k}\cdot[{\bm X}^\prime_1\times{\bm X}^\prime_2])e^{i{\bm k}\cdot{\bm X}_{21}}&=&2{\bm k}^2\int d\sigma_1d\sigma_2(\delta{\bm X}_1\cdot[{\bm X}^\prime_1\times{\bm X}^\prime_2])e^{i{\bm k}\cdot{\bm X}_{21}},
\end{eqnarray}
\end{widetext}It is suitable to represent the energy $E=E_{\rm even}+E_{\rm odd}$ and its variation in the form of the sum of parity-even and parity-odd contributions characterized in Eq.~(\ref{Ffin}), respectively,  by the even ($\propto I_1$) and odd ($\propto I_2$) powers of the parameter $\alpha$.  Using Eq.~(\ref{variate}) one obtains
\begin{widetext}
\begin{eqnarray}\label{deltaF}
\delta E_{\rm even}&=&\frac{\Phi^2_0}{32\pi^2\lambda^2_L}\left(1-\alpha^2\right)\int
\frac{d\sigma_1 d\sigma_2}{|{\bm X}_{21}|}{\bm X}_{21}\cdot\left([{\bm X}^\prime_2[{\bm X}^\prime_1\times\delta{\bm X}_1]]-[{\bm X}^\prime_1[{\bm X}^\prime_2\times\delta{\bm X}_2]]\right)\times\nonumber\\&&\frac{\partial}{\partial|{\bm X}_{21}|}\left[\frac{e^{-|{\bm X}_{21}|\sqrt{1-\alpha^2}/\lambda_L}}{|{\bm X}_{21}|}\left(\cos\frac{\alpha|{\bm X}_{21}|}{\lambda_L}+\frac{\alpha}{\sqrt{1-\alpha^2}}\sin\frac{\alpha|{\bm X}_{21}|}{\lambda_L}\right)\right],\nonumber\\
\delta E_{\rm odd}&=&\frac{\Phi^2_0\alpha(1-\alpha^2)}{8\pi^2\lambda^3_L}\int d\sigma_1d\sigma_2\frac{([{\bm X}^\prime_1\times{\bm X}^\prime_2]\cdot\delta{\bm X}_1)}{|{\bm X_{21}}|}\left(\cos\frac{\alpha|{\bm X}_{21}|}{\lambda_L}+\frac{2\alpha^2-1}{2\alpha\sqrt{1-\alpha^2}}\sin\frac{\alpha|{\bm X}_{21}|}{\lambda_L}\right)
\times\nonumber\\&&e^{-|{\bm X}_{21}|\sqrt{1-\alpha^2}/\lambda_L}.
\end{eqnarray}\end{widetext}Our goal is to obtain the local form which is exact in $\alpha$ but to the first order in the product $\kappa^2\lambda^2_L\ll1$, where $\kappa^2={\bm X}^{\prime\prime2}$ is the square   of the contour curvature. See Eq.~(\ref{frenet}). This can be done upon using the expansion
\begin{eqnarray}\label{expan}
{\bm X}(\sigma_2)&=&{\bm X}(\sigma_1)+\frac{z}{1!}{\bm X}^\prime(\sigma_1)+\frac{z^2}{2!}{\bm X}^{\prime\prime}(\sigma_1)+\nonumber\\&&\frac{z^3}{3!}{\bm X}^{\prime\prime\prime}(\sigma_1)+\ldots,
\end{eqnarray}where $z=\sigma_2-\sigma_1$. The integration in Eq.~(\ref{deltaF}) can be represented in the form $\int d\sigma_1d\sigma_2\approx\int d\sigma\int_{-\infty}^\infty dz$.  The limits of integration over $z$ can be set to $\pm\infty$ in view of exponential damping of the gauge field profile at large distances. When integrating over $z$ one should use the integral
\begin{eqnarray}\label{Int}
\int_0^\infty z^n\left[
                   \begin{array}{c}
                     \cos\frac{\alpha z}{\lambda_L}\\
                     \sin\frac{\alpha z}{\lambda_L}\\
                   \end{array}
                 \right]e^{-z\sqrt{1-\alpha^2}/\lambda_L} dz&=&\nonumber\\n!\lambda_L^{n+1}\left[
                                              \begin{array}{c}
                                                \cos\frac{(n+1)\alpha}{\sqrt{1-\alpha^2}} \\
                                                \sin\frac{(n+1)\alpha}{\sqrt{1-\alpha^2}} \\
                                              \end{array}
                                            \right].
\end{eqnarray}

First, let us evaluate the parity-even contribution $E_{\rm even}$. Using Eq.~(\ref{deltaF}) and the expansion Eq.~(\ref{expan}) one gets
\begin{eqnarray}
\label{deltaFeven}
\delta E_{\rm even}&=&\frac{\Phi^2_0\left(1-\alpha^2\right)}{16\pi^2\lambda^2_L}\int_0^\infty dzz\frac{\partial}{\partial z}\left[\frac{e^{-z\sqrt{1-\alpha^2}/\lambda_L}}{z}\times\right.\nonumber\\&&\left.\left(\cos\frac{\alpha z}{\lambda_L}+\frac{\alpha}{\sqrt{1-\alpha^2}}\sin\frac{\alpha z}{\lambda_L}\right)\right]\times\nonumber\\&&
\int d\sigma({\bm X}^{\prime\prime}\cdot\delta{\bm X})=\nonumber\\&&\frac{\Phi^2_0G_1(\alpha)}{16\pi^2\lambda^2_L}\int d\sigma({\bm X}^{\prime\prime}\cdot\delta{\bm X}),
\end{eqnarray}where
\begin{eqnarray}\label{G1}
G_1(\alpha)&=&(1-\alpha^2)\left(\ln\frac{\lambda_L}{\xi\sqrt{1-\alpha^2}}-\frac{\alpha}{\sqrt{1-\alpha^2}}\times\right.\nonumber\\&&\left.
\arcsin\alpha+\frac{2\alpha^2-1}{\sqrt{1-\alpha^2}}\cos\frac{\alpha}{\sqrt{1-\alpha^2}}-\right.\nonumber\\&&
\left.2\alpha\sin\frac{\alpha}{\sqrt{1-\alpha^2}}\right)\approx(1-\alpha^2)\ln\frac{\lambda_L}{\xi}.
\end{eqnarray}
Here, the divergence  at short distances is regularized as is typical to the London limit. Recall that the modulus of the Higgs field $n^{1/2}_0$ is assumed to be constant everywhere except the vortex core where it goes to zero at the distance $\sim\xi$ ($1/m_H$). The answer is valid with the logarithmic accuracy, hence the approximate equality in Eq.~(\ref{G1}). However, it cannot be valid at $\alpha$ too close to unity. Nevertheless, one can estimate the limit of validity of the large logarithm approximation. The estimate from Eq.~(\ref{G1}) gives
\begin{equation}\label{eps}
1-|\alpha|\gg\left(2\ln\frac{\lambda_L}{\xi}\right)^{-2}.
\end{equation}Taking $\ln\frac{\lambda_l}{\xi}\sim10$ one obtains $1-|\alpha|\gg1/400$. Note also that the corrections $O(\kappa^2\lambda^2_L)$ to the parity-even contribution do not include large logarithm and can be neglected \cite{kozh10}. In fact, the above inequality states the limit of validity of the London limit adopted in the present work.

\begin{figure}
\includegraphics[width=11cm]{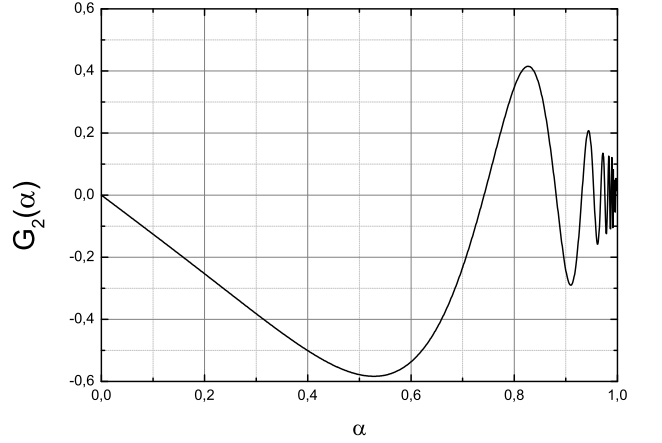}
\caption{\label{Galpha}The function describing the coupling strength dependence of the space parity violating term in the vortex equation of motion.}
\end{figure}

The variation of the parity-odd contribution is obtained in the same manner:
\begin{equation}\label{deltaFoddloc}
\delta E_{\rm odd}=\frac{\Phi^2_0\lambda_L G_2(\alpha)}{8\pi^2}\int d\sigma\kappa^2([{\bm X}^\prime\times{\bm X}^{\prime\prime\prime}]\cdot\delta{\bm X}),
\end{equation}where the function $G_2(\alpha)$ looks as follows:
\begin{eqnarray}\label{G2}
G_2(\alpha)&=&\alpha(1-\alpha^2)\left[\frac{1}{4}\cos\frac{4\alpha}{\sqrt{1-\alpha^2}}+\right.\nonumber\\&&\left.
\frac{2\alpha^2-1}{8\alpha\sqrt{1-\alpha^2}}\sin\frac{4\alpha}{\sqrt{1-\alpha^2}}+\right.\nonumber\\&&\left.
\left(1-\frac{2\alpha^2-1}{2\sqrt{1-\alpha^2}}\right)\cos\frac{5\alpha}{\sqrt{1-\alpha^2}}+\right.\nonumber\\
&&\left.\left(\alpha+\frac{2\alpha^2-1}{2\alpha\sqrt{1-\alpha^2}}\right)\sin\frac{5\alpha}{\sqrt{1-\alpha^2}}\right].
\end{eqnarray}It is the odd function, $G_2(-\alpha)=-G_2(\alpha)$. The plot of this function at $\alpha\geq0$ is shown in Fig.~\ref{Galpha}. Its behavior at $\alpha\ll1$ is $G_2(\alpha)=-5\alpha/4$, and $G(\alpha)$ goes to 0 non-analytically when $\alpha\to1$. One should have in mind that $\alpha=1$ is the border of the vortex state stability of the present model, i.e. the energy is positive at $\alpha<1$. See Ref.~\cite{kharz20}. One can see that $E$ being proportional to $1-\alpha^2$ vanishes everywhere at $\alpha=1$. Since the effective London penetration depth is $\lambda_{\rm eff}=\lambda_L/\sqrt{1-\alpha^2}$, this means the full penetration of the magnetic field at $\alpha\to1$ and demands going beyond the London limit to take into account the Higgs field profile. See Ref.~\cite{kharz20}.
--------------------------------------------------------------------------------

\section{Equation of motion and the static solution}\label{eqmo}
~

To obtain the equation of the vortex motion one should vary the effective action over ${\bm X}(\sigma,t)$. All concerning the variation $\delta E$ is done in the previous section. The remaining piece is obtained by varying the $a_0$ term in Eq.~(\ref{Seff}) with the help of Eq.~(\ref{amuk}) to give \cite{kozh10}
\begin{eqnarray}\label{deltaa0}
\delta\int d^4xa_0&=&-\frac{\Phi_0}{c}\int dtd\sigma([\dot{{\bm X}}\times{\bm X}^\prime]\cdot\delta{\bm X})
\end{eqnarray}where the overdot means the derivative over time variable. The equation of the curved vortex motion looks as follows:
\begin{eqnarray}\label{eqmo1}
[\dot{{\bm X}}\times{\bm X}^\prime]&=&\frac{\hbar G_1(\alpha)}{2m}{\bm X}^{\prime\prime}+\nonumber\\&&
\frac{\hbar\lambda_L^3G_2(\alpha)}{m}\kappa^2[{\bm X}^\prime\times{\bm X}^{\prime\prime\prime}].
\end{eqnarray}As is evident from Eq.~(\ref{eqmo1}) and Fig.~\ref{Galpha}, there is a discrete series of parameters for which $G_2(\alpha_n)=0$, so that the parity-odd effects in the vortex motion disappear at $\alpha=\alpha_n\not=0$. In the following treatment we will assume that $\alpha\not=\alpha_n$.
Introducing the vector ${\bm W}=[\dot{{\bm X}}\times{\bm X}^\prime]$ one finds from Eq.~(\ref{eqmo1}):
\begin{eqnarray}\label{Wnb}
W_n&\equiv&({\bm W}\cdot{\bm n})=\kappa(a-d\kappa^2\tau),\nonumber\\
W_b&\equiv&({\bm W}\cdot{\bm b})=d\kappa^2\kappa^\prime,\nonumber\\
W_\parallel&\equiv&({\bm W}\cdot{\bm X}^\prime)=0,
\end{eqnarray}where
\begin{eqnarray}\label{ab}
a&=&\frac{\hbar G_1(\alpha)}{2m},\nonumber\\
d&=&\frac{\hbar\lambda_L^3 G_2(\alpha)}{m}.
\end{eqnarray}Eq.~(\ref{Wnb}) are obtained upon taking into account the pure geometric Frenet-Serre equations
\begin{eqnarray}
{\bm X}^{\prime\prime}&=&\kappa{\bm n},\nonumber\\
{\bm n}^\prime&=&-\kappa{\bm X}^\prime+\tau{\bm b},\nonumber\\
{\bm b}^\prime&=&-\tau{\bm n}, \label{frenet}
\end{eqnarray}
where $\tau$ stands for the torsion of the contour, and ${\bm n}$, ${\bm b}$ are the vectors of normal and bi-normal, respectively. The  vectors $({\bm n},{\bm b},{\bm X}^\prime)$ comprise the right triple of the unit orthogonal vectors, so that ${\bm X}^\prime=[{\bm n}\times{\bm b}]$ (and  similar relations obtained by the cyclic permutation).

The longitudinal component of velocity $v_\parallel=(\dot{{\bm X}}\cdot{\bm X}^\prime)$ cannot be found from Eq.~(\ref{eqmo1}). One can find it from the requirement that the gauge condition ${\bm X}^{\prime2}=1$ should be satisfied during evolution \cite{kharz18}. This results in the relation $(\dot{{\bm X}^\prime}\cdot{\bm X}^\prime)=0$. Representing the velocity in the form
$$\dot{{\bm X}}=W_n{\bm b}-W_b{\bm n}+v_\parallel{\bm X}^\prime$$ and taking derivative of this expression over $\sigma$ one obtains that
\begin{eqnarray}\label{dotXprim}
\dot{{\bm X}^\prime}&=&(-W^\prime_b-\tau W_n+\kappa v_\parallel){\bm n}+(W^\prime_n-\tau W_b){\bm b}+\nonumber\\&&
(v^\prime_\parallel+\kappa W_b){\bm X}^\prime,
\end{eqnarray}resulting, in particular, in the expression
\begin{equation}\label{v||}
v_\parallel^\prime=-\kappa W_b.
\end{equation}
Taking into account Eqs.~(\ref{Wnb}) and (\ref{v||}) one can see that the condition of preserving the gauge constraint  ${\bm X}^{\prime2}=1$  is \begin{equation}\label{v||1}
v_\parallel=\frac{d}{4}(\kappa^4_0-\kappa^4),
\end{equation}
where $\kappa_0$ is the constant of integration. Then the expression for the gauge vortex velocity can be represented in following equivalent forms:
\begin{eqnarray}\label{veloc}
\dot{{\bm X}}&=&\kappa(a-d\kappa^2\tau){\bm b}-\frac{d}{3}(\kappa^3)^\prime{\bm n}+\frac{d}{4}(\kappa^4_0-\kappa^4){\bm X}^\prime,\nonumber\\
\dot{{\bm X}}&=&a[{\bm X}^\prime\times{\bm X}^{\prime\prime}]-d\kappa^2\left({\bm X}^{\prime\prime\prime}+\frac{5}{4}\kappa^2{\bm X}^\prime\right)+\nonumber\\&&\frac{d\kappa^4_0}{4}{\bm X}^\prime.
\end{eqnarray}
The equivalence is verified with the help of Eq.~(\ref{frenet}).

The static vortex contour should satisfy the condition $\dot{{\bm X}}=0$, hence $\kappa=\kappa_0$ and
\begin{equation}\label{stat1}
(a-d\kappa^2\tau)=0.
\end{equation}The static torsion $\tau_0$ is related with the static curvature $\kappa_0$:
\begin{eqnarray}\label{statsol}
\tau&=&\tau_0=\frac{a}{d\kappa^2_0}=\frac{G_1(\alpha)}{2\lambda^3_L\kappa^2_0G_2(\alpha)}.
\end{eqnarray}The contour with constant curvature $\kappa$ and torsion $\tau$  is a helix \cite{dubrovin}.  Because $G_1>0$, the sign of the torsion $\tau$, that is, is the helix right or left, depends on the sign of the function $G_2(\alpha)$ plotted in Fig.~\ref{Galpha}. To be specific, let us choose the $z-$oriented helix parameterized with the radius $R$ and step $h$ (do not confuse with Planck constant) of the winding:
\begin{equation}\label{helix}
{\bm X}_0(\sigma)=R\left({\bm e}_x\cos\frac{\sigma}{l}+{\bm e}_y\cos\frac{\sigma}{l}\right)+\frac{h\sigma}{2\pi l}{\bm e}_z,
\end{equation}where
\begin{equation}\label{l}
l=\sqrt{R^2+\left(\frac{h}{2\pi}\right)^2}.
\end{equation}Then the basic contour characteristic are
\begin{eqnarray}
\label{contchar}
\kappa&=&R/l^2,\nonumber\\
\tau&=&h/2\pi l^2.
\end{eqnarray}The relation Eq.~(\ref{statsol}) reduces to the relation between $R$ and $h$ which, upon introducing the ratio $x=h/2\pi R$, reads
\begin{equation}\label{Rh}
\frac{(1+x^2)^3}{x}=2\left(\frac{\lambda_L}{R}\right)^3\frac{G_2(\alpha)}{G_1(\alpha)}.
\end{equation}To be specific, let us take the parameter $\alpha$ such that $G_2(\alpha)>0$. One should conform Eq.~(\ref{Rh}) with the condition of validity of the present treatment, $\kappa^2\lambda^2_L\ll1$. Taking into account Eqs.~(\ref{G1}), (\ref{G2}), and (\ref{contchar}) one obtains that
\begin{equation}\label{condtreat}
x\gg\frac{G_1(\alpha)}{2G_2(\alpha)}.
\end{equation}Using numerical values of $G_{1,2}(\alpha)$ at $\alpha\not=\alpha_n$ [recall that $G_2(\alpha_n)=0$] one can convince that $x\gg1$. Qualitatively, the overall treatment  is valid for helices with the step $h$ much greater than the radius of winding $R$.

\section{Small oscillations around static shape}
\label{oscill}
~

Let us turn to the dynamical treatment of the problem and consider the small oscillations around the static contour shape found in the previous section. To this end one should take ${\bm X}(\sigma,t)={\bm X}_0(\sigma)+{\bm\xi}(\sigma,t)$, where ${\xi}$ is the small deviation from the static contour ${\bm X}_0$, and substitute this to the equation of motion keeping the terms up to the first order in ${\xi}$. One obtains the equation
\begin{eqnarray}\label{xidot}
[\dot{{\bm\xi}}\times{\bm X}^\prime_0]&=&a{\bm\xi}^{\prime\prime}-2a({\bm n}_0\cdot{\bm\xi}^{\prime\prime}){\bm b}_0+\nonumber\\&&d\kappa^2_0\left([{\bm X}^\prime_0\times{\bm\xi}^{\prime\prime\prime}]+[{\bm\xi}^\prime\times{\bm X}^{\prime\prime\prime}_0]\right).
\end{eqnarray}
Hereafter the quantities with the index 0 refer to the unperturbed contour ${\bm X}_0(\sigma)$, with the corresponding vectors of normal ${\bm n}_0$, bi-normal ${\bm b}_0$, and tangent ${\bm X}^\prime_0$, and the curvature $\kappa_0$ and torsion $\tau_0$. These three vectors can be considered as the local coordinate frame. Taking the scalar products of Eq.~(\ref{xidot}) by, respectively, ${\bm X}^\prime_0$, ${\bm n}_0$, and ${\bm b}_0$ one obtains the relation $a({\bm X}^\prime_0\cdot{\bm\xi}^\prime)^\prime=0$ which is integrated to give the constraint
\begin{equation}\label{constr}
({\bm X}^\prime_0\cdot{\bm\xi}^\prime)=0,
\end{equation}and the following dynamical equations
\begin{eqnarray}\label{xibndot}
\dot{\xi}_b&\equiv&({\bm b}_0\cdot\dot{{\bm\xi}})=-a({\bm n}_0\cdot{\bm\xi}^{\prime\prime})-d\kappa^2_0\left[({\bm b}_0\cdot{\bm\xi}^{\prime\prime\prime})+\right.\nonumber\\&&\left.\kappa^2_0({\bm b}_0\cdot{\bm\xi}^\prime)\right],\nonumber\\
\dot{\xi}_n&\equiv&({\bm n}_0\cdot\dot{{\bm\xi}})=-a({\bm b}_0\cdot{\bm\xi}^{\prime\prime})-d\kappa^2_0\left[({\bm n}_0\cdot{\bm\xi}^{\prime\prime\prime})+\right.\nonumber\\&&\left.\kappa^2_0({\bm n}_0\cdot{\bm\xi}^\prime)\right].
\end{eqnarray}Note that Eq.~(\ref{constr}) provides that the relations ${\bm X}^{\prime2}=1$ and $({\bm X}^\prime\cdot[{\bm n}\times{\bm b}])=1$ are preserved  in the course of the contour evolution up to the first order in ${\bm\xi}$. Since
\begin{equation}\label{vecxi}
{\bm\xi}=\xi_n{\bm n}_0+\xi_b{\bm b}_0+\xi_\parallel{\bm X}^\prime_0,
\end{equation}Eq.~(\ref{constr}) results in the relation
\begin{equation}\label{xit}
\xi^\prime_\parallel=\kappa_0\xi_n
\end{equation}
which shows that the only dynamical quantities in the gauge vortex dynamics are the locally transverse quantities $\xi_n$ and $\xi_b$. By taking the derivatives over $\sigma$ and using  Eq.~(\ref{frenet}) one can obtain the scalar products of the above by the unit vectors ${\bm n}_0$, ${\bm b}_0$, and ${\bm X}^\prime_0$ resulting in the equations
\begin{eqnarray}\label{xinbdyn}
\dot{\xi}_n&=&\frac{a}{\tau_0}\left[-\xi_n^{\prime\prime\prime}+2\tau_0\xi_b^{\prime\prime}+\left(\tau^2_0-\kappa^2_0\right)\xi_n^\prime\right],\nonumber\\
\dot{\xi}_b&=&\frac{a}{\tau_0}\left[-\xi_b^{\prime\prime\prime}-4\tau_0\xi_n^{\prime\prime}+\left(5\tau^2_0-\kappa^2_0\right)\xi_b^\prime+
\right.\nonumber\\&&\left. 2\tau_0\left(\tau^2_0-2\kappa^2_0\right)\xi_n\right].
\end{eqnarray}The plane wave solution $${\bm\xi}(\sigma,t)=(C_n,C_b,C_\parallel)e^{-i\omega t+ik\sigma}$$ gives the dispersion law
\begin{eqnarray}\label{omegapm}
\omega_\pm(k)&=&\frac{a}{\tau_0}\left[-k^3-(3\tau^2_0-\kappa^2_0)k\pm\right.\nonumber\\&&\left.
2\sqrt{2k^2\tau^2_0(k^2+\tau^2_0-\kappa^2_0)}\right].
\end{eqnarray}
The potential instability of small oscillations could arise when $\tau^2_0-\kappa^2_0<0$. However, because of the relation
\begin{eqnarray*}
\tau^2_0-\kappa^2_0&=&\kappa^2_0\left\{\left[\frac{G_1(\alpha)/2G_2(\alpha)}{(\kappa_0\lambda_L)^3}\right]^2-1\right\},
\end{eqnarray*}and the inequality $\kappa_0\lambda_L\ll1$, one can see that $\tau^2_0-\kappa^2_0>0$ in the domain of applicability of the present treatment, so the instability domain cannot be reached. In fact, the stronger inequality
$\tau_0\gg\kappa_0$ takes place. See the discussion following inequality (\ref{condtreat}).

The relations between the Fourier amplitudes are
\begin{eqnarray}\label{Cnbt}
C_n^{(\pm)}&=&\frac{ikC_b^{(\pm)}}{\tau_0\mp\sqrt{2(k^2+\tau^2_0-\kappa^2_0)}},\nonumber\\
C_\parallel^{(\pm)}&=&-\frac{i\kappa_0}{k}C_n^{(\pm)}=\frac{\kappa_0C_b^{(\pm)}}{\tau_0\mp\sqrt{2(k^2+\tau^2_0-\kappa^2_0)}}.
\end{eqnarray}Recall that the second line of the above relations is the consequence of the constraint Eq.~(\ref{xit}). The general solution is represented as the sum over modes:
\begin{widetext}
\begin{eqnarray}\label{xigensol}
{\bm\xi}(\sigma,t)&=&\int\frac{dk}{2\pi}\left\{C_b^{(+)}(k)\left[{\bm b}_0+\frac{ik{\bm n}_0+\kappa_0{\bm X}^\prime_0}{\tau_0-\sqrt{2(k^2+\tau^2_0-\kappa^2_0)}}\right]e^{-i\omega_+(k)t}+C_b^{(-)}(k)\left[{\bm b}_0+\frac{ik{\bm n}_0+\kappa_0{\bm X}^\prime_0}{\tau_0+\sqrt{2(k^2+\tau^2_0-\kappa^2_0)}}\right]\times\right.\nonumber\\&&\left.e^{-i\omega_-(k)t}\right\}e^{ik\sigma}+{\rm c.c.}.
\end{eqnarray}
\end{widetext}A visible singularity  for the mode with $\omega=\omega_+$ (in case of $\tau_0>0$) and one with $\omega=\omega_-$ (in case of $\tau_0<0$) arising at the wave number $k=\pm\sqrt{\kappa^2_0-\tau^2_0/2}$, in view of Eq.~(\ref{Cnbt}), means the vanishing of the corresponding $b$-component of the vortex displacement at this wave number. But, in fact, this singularity cannot be encountered because $\tau_0\gg\kappa_0$ in the present treatment. It is important that the dynamics of small oscillations is expressed solely through the local contour variables ${\bm n}_0(\sigma)$ (curvature), $\tau_0(\sigma)$ (torsion), and ${\bm X}^\prime_0(\sigma)$ (tangent vector). One can express ${\bm\xi}(\sigma,t)$ in terms of the global unit vectors ${\bm e}_{x,y,z}$ for the specific orientation of the helix, as, for example, specified by Eq.~(\ref{helix}).

The dynamics of curvature $\kappa(t)$ and torsion $\tau(t)$ can be obtained, in particular, along the lines presented in Refs.~\cite{kozh15,kozh16}. The  equations of motion of curvature and torsion read
\begin{eqnarray}\label{kappataueqmo}
\dot{\kappa}&=&(-W_b^\prime-\tau W_n+\kappa v_\parallel)^\prime-\tau(W^\prime_n-\tau W_b),\nonumber\\
\dot{\tau}&=&\left\{\frac{1}{\kappa}\left[(W^\prime_n-\tau W_b)^\prime+\tau(-W_b^\prime-\tau W_n+\kappa v_\parallel)\right]\right\}^\prime+\nonumber\\&&
\kappa(W^\prime_n-\tau W_b).
\end{eqnarray}The longitudinal component of the velocity $v_\parallel=(\dot{{\bm X}}\cdot{\bm X}^\prime)$ is given by Eqs.~(\ref{v||}), (\ref{v||1}).  After expanding the system Eq.~(\ref{kappataueqmo}) near the static solution Eq.~(\ref{statsol}), $\kappa=\kappa_0+\delta\kappa$ and $\tau=\tau_0+\delta\tau$, one arrives at the equations for small deviations:
\begin{eqnarray}\label{dotdelkappatau}
\frac{\partial\delta\kappa}{\partial t}&=&\frac{a}{\tau_0}\left[-\delta\kappa^{\prime\prime\prime}+(5\tau^2_0-\kappa^2_0)\delta\kappa^\prime+2\kappa_0\tau_0\delta\tau^\prime\right],\nonumber\\
\frac{\partial\delta\tau}{\partial t}&=&\frac{a}{\tau_0}\left[-\delta\tau^{\prime\prime\prime}+(\tau^2_0-\kappa^2_0)\delta\tau^\prime-
\frac{4\tau_0}{\kappa_0}\delta\kappa^{\prime\prime\prime}-\right.\nonumber\\&&\left.\frac{2\tau_0}{\kappa_0}\delta\kappa^\prime\right].
\end{eqnarray}The plane wave solutions,
\begin{eqnarray}
\delta\kappa(\sigma,t)&=&C_\kappa e^{-i\omega t+ik\sigma},\nonumber\\
\delta\tau(\sigma,t)&=&C_\tau e^{-i\omega t+ik\sigma},\end{eqnarray} result in the dispersion law which coincides with Eq.~(\ref{omegapm}), and the relationship between the Fourier amplitudes:
$$C^{(\pm)}_\kappa=-\frac{\kappa_0C^{(\pm)}_\tau}{\tau_0\pm\sqrt{2(k^2+\tau^2_0-\kappa^2_0)}}.$$The longitudinal component of velocity,
$\delta v_\parallel=-a\kappa_0\delta\kappa/\tau_0$, is determined by the dynamics of the curvature and has not a proper dynamical meaning. As usual, the general solution for $\kappa(t)$ and $\tau(t)$ is represented by the sum over these modes.

\section{Discussion}\label{discus}
~

The main goal of the present work is to consider the shape and dynamics of the curved  nonrelativistic gauge vortex string upon taking into account the effects of the space parity breaking environment. So additional terms in the vortex equation of motion arising due to the exchange of the Bogolyubov-Anderson excitations between distant sectors of the string \cite{kozh10,kozh15,kozh16} were neglected. Using the results of the cited works one can show that the criterion of this is the following inequality:
\begin{equation}\label{neglect}
(\xi\kappa_0)^2\ln\frac{\lambda_L}{\xi}\ln\left(\frac{\lambda_L\hbar}{mc\xi^2}\right)\ll1.
\end{equation}Despite the fact that product of logarithms in the above inequality can be large, the inequality is satisfied because   of the inequalities  $\kappa_0\lambda_L\ll1$ (thin string approximation adopted in the present work) and $\xi\ll\lambda_L$ (the London limit).

Let us compare the equation of motion (\ref{eqmo1}) describing the dynamics in case of the fixed B-type helicity with the analogous equations which take place in the media with the chiral imbalance \cite{kozh99,kozh15,kozh16,kharz18} corresponding to fixing the A-type one. In these works the term in the equation of the gauge vortex motion induced by quantum anomaly, $\propto[{\bm X}^\prime\times{\bm X}^{\prime\prime\prime}]$, appears to be multiplied by the chemical potential $\mu_F$ characterizing the chiral imbalance. Hence, the chiral imbalanced environment exerts the gauge vortex string motion at any  $\mu_F\not=0$. The case of the gauge vortex string in the parity-odd media, as, for example that suggested in Ref.~\cite{kharz20,babaev20}, which shape and dynamics in the curved situation is considered in the present work, is completely different. First, the influence  of the parity-odd environment on the vortex equation of motion is governed by the function $G_2(\alpha)$, Eq.~(\ref{G2}), with infinite discrete series of zeros $\alpha_n$, see Fig.~\ref{Galpha} where $\alpha$ given by Eq.~(\ref{alpha}) characterizes the strength of the parity breaking term (Lifshitz invariant) in the effective action. So, at $\alpha=\alpha_n$ the vortex does not feel the parity-breaking environment. Second, the term $\propto[{\bm X}^\prime\times{\bm X}^{\prime\prime\prime}]$ in Eq.~(\ref{eqmo1})  enters with the factor $\kappa^2$ (the contour curvature squared) which, formally, is the result of fixing the $B$-type helicity in the Feynman path integral. Such factor is absent in case of fixing the $A$-type helicity. This is the reason of appearance of the relation Eq.~(\ref{statsol}) between the curvature $\kappa_0$ and and torsion $\tau_0$ of the static helical contour. The small oscillations around this static helical contour given by Eq.~(\ref{xigensol}) demonstrate nontrivial polarization properties which depend on the wave number $k$.

It is interesting to compare the second expression for the gauge vortex velocity in Eq.~(\ref{veloc}) with its analog in case of fixing the $A$-type helicity (\ref{haaver}). In this situation, the parity-odd contribution to velocity $\dot{{\bm X}}$, up to constant multipliers, is analogous to Eq.~(\ref{eqmo1}), but appears without the multiple $\kappa^2={\bm X}^{\prime\prime2}$ \cite{kozh99,kozh15,kozh16}. Repeating the derivation that results in the second line of Eq.~(\ref{veloc}), one arrives at the expression
$$\dot{{\bm X}}=a[{\bm X}^\prime\times{\bm X}^{\prime\prime}]-d\left({\bm X}^{\prime\prime\prime}+\frac{3}{2}\kappa^2{\bm X}^\prime\right)+\frac{d\kappa^2_0}{2}{\bm X}^\prime,$$where $d$ is proportional to the chemical potential characterizing the chiral imbalance. Taking integration constant $\kappa_0=0$ and setting $a=1$ one gets the nonlinear equation \cite{kharz18} studied in hydrodynamics \cite{fuk} which, using the Hasimoto transformation \cite{hasim},
\begin{equation}\label{has}
\psi=\kappa e^{i\int^\sigma_0\tau d\sigma},
\end{equation}
can be mapped into the integrable Hirota equation \cite{hirota},
\begin{equation}\label{hir}
i\dot{\psi}+\psi^{\prime\prime}+\frac{1}{2}|\psi|^2\psi+id\left[\psi^{\prime\prime\prime}+\frac{3}{2}|\psi|^2\psi\right]=0.
\end{equation}
(See Ref.~\cite{kharz18} for the detailed studies of the nonlinear soliton dynamics and small oscillations around specific contour shapes.) The analogous mapping can be made for nonlinear equation (\ref{veloc}) to obtain
\begin{equation}\label{has1}
i\dot{\psi}+\psi^{\prime\prime}+\frac{1}{2}|\psi|^2\psi+id\left[(|\psi|^2\psi^\prime)^{\prime\prime}+\frac{5}{4}|\psi|^4\psi^\prime\right]=0,
\end{equation}where the time variable has been re-scaled to set $a=1$ in Eq.~(\ref{veloc}), and $\kappa_0$ was set to zero. Is this equation integrable like Eq.~(\ref{hir}) or not is an open question.

The study was carried out within the framework of the state contract
of the Sobolev Institute of Mathematics (project no. 0314-2019-0021).

\end{document}